# POLARIZATION IN A MUON COLLIDER

D. Cline, UCLA, Los Angeles, USA
B. Norum, University of Virginia, Charlottesville, USA
R. Rossmanith, DESY, Hamburg, Germany

## 1 INTRODUCTION

In various recent publications new concepts for a high energy muon collider was presented [1,2,3]. It is claimed that a muon collider could be ideal for high energy, high luminosity lepton collisions. In this paper the possibility of obtaining polarized muon beams is discussed.

Muons are generated in a pion decay. The pion decays via the weak interaction into a muon and a muon neutrino (fig. 1). The muons are born polarized when the pions are at rest.

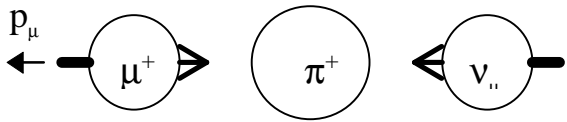

Figure 1. The muon is polarized opposite to its direction of motion.

The rest mass of the muon is 105.66 MeV and therefore about 207 times larger than the rest mass of the electron. The pion rest mass is 139.6 MeV. The anomalous magnetic moment of the muon is $1.166 \times 10^{-3}$ similar to the electron anomalous magnetic moment of $1.16 \times 10^{-3}$.

During acceleration and storage the integer resonances occur when (n = integer)

$$\left(\frac{g-2}{2}\right)\gamma = n$$

The distance between two integer resonances is 94.62 GeV compared to ca. 440 MeV with electrons.

The figures clearly show that the muon spin is extremely insensitive to magnetic fields. This has a positive aspect. Polarization is far more stable than it is in the case of electrons and protons. However, there is also a negative aspect: in the relativistic limit an integrated field of 492 Tesla.m is required to rotate the muon spin by 90 degrees. In comparison, 2.3 T.m are required to rotate the spin of an electron by 90 degrees.
The clear advantage is that even at an end energy of 2 TeV the spin tune is only 21.13 and comparable to a ca. 10 GeV electron storage ring.

## 2 THE GENERATION OF POLARIZED MUONS

In the muon collider the muons are produced in a so-called decay channel by moving pions. When the pion moves the momenta of the muon and the pion add or subtract ('forward' or 'backward' muon) [4].

$$p_\mu = \frac{(m_\pi^2 - m_\mu^2)(p_\pi^2 + m_\pi^2 c^2)^{1/2} \pm p_\pi (m_\pi^2 + m_\mu^2)}{2 m_\pi^2}$$

Forward and backward muons have different spin directions. The plus and the minus sign describes the two possibilities.

When it is assumed that the pions are monochromatic and have a momentum of 200 MeV/c, muons with a momentum of 209 MeV/c (forward muon) or 105 MeV/c (backward muon) are produced. The muon beam is unpolarized. Polarization is obtained by energy selection. The energy of the produced muons lies between 0 to 3 GeV and can only be captured by a special linac which unifies the energies of the muons. The momentum of the forward and backward muons as a function of the momentum of the pion is shown in fig. 2.

In order to obtain polarized muons, the high energy muons have to be selected (fig. 2). This reduces the maximum polarized current and therefore the luminosity. A polarization of 60% would reduce the polarized luminosity by about a factor of 20.

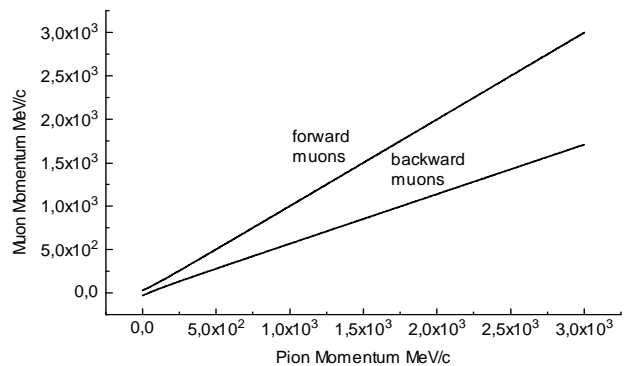

Fig. 2 Momenta of backward and forward muons as a function of pion momentum

## 3 DEPOLARIZATION IN THE COOLING ELEMENT

The muon beam has to be cooled by ionizarion cooling [2]. In order to estimate the degree of depolarization during ionization cooling, it is assumed that the binding energy of the electrons is small compared to the muon energy: the electrons are considered as free and unpolarized particles.

In the nonrelativistic approach, the probability Q that a muon is scattered into the angle $\theta$ is given by [5]

$$Q(\varepsilon,\theta) = \frac{1+\varepsilon}{2} - \varepsilon \frac{m_e^2}{m_\mu^2} \beta^4 \cdot [\sin^2(\theta/2) - \sin^4(\theta/2) - \sin^6(\theta/2)]$$

$\theta$ is the center of momentum angle. $\varepsilon$ is +1 when the spin remains the same after scattering. $\varepsilon$ is -1 when the spin is flipped. $m_e$ is the electron rest mass and $m_\mu$ is the rest mass of the muon. $\beta$ is v/c in the usual definition.

The spin flip probability is proportional to the fractional energy loss. The proportionality constant is

$$\left(\frac{m_e}{m_\mu}\right)\beta^2$$

It becomes clear that depolarization is almost negligible when $\beta$ is small

The depolarization can be described by the following formula

$$P = P_0 \exp[-x/a]$$

P is polarization, x the length of the particle trajectory in the material. a is the decay length of the polarization. For Be and the given energy [6] a is about 200 m.

## 4 SYNCHROTRON SPIN MATCHING FOR HORIZONTALLY POLARIZED BEAMS IN RECIRCULATING LINACS

At low energies it seems to be difficult to rotate the spin into the vertical direction. It seems to be easier to keep the polarization in the horizontal plane as long as possible and to rotate it at higher energies when the bending fields become higher.

Recirculatig linacs are ideal for this case. During acceleration particles at the trailing and the leading edge of the bunch gain different energies. When the recirculating arc has an appropriate momentum compaction factor particles change the position in the bunch from revolution to revolution. We call this technique synchrotron spin matching [7].

In the case of synchrotron spin matching the number of revolutions, the momentum compaction factor and the acceleration voltage all have to be matched appropriately. The matching condition is

$$\int \left(\frac{g-2}{2}\right) \Delta\gamma(t)dt = 0$$

in the relativistic limit where $\Delta\gamma$ is the energy deviation. Is this condition fulfilled the beam remains perfectly polarized.

## 5 SPIN HANDLING IN THE STORAGE RING

A general type of spin rotator in the main ring is shown in fig. 3. It is assumed that the field strength of the bending magnets will be ca. 10 T and that the magnets are about 10 m long. As a result, a 30 m long magnet assembly consisting of 3 magnets is able to rotate the spin by 45 degrees.

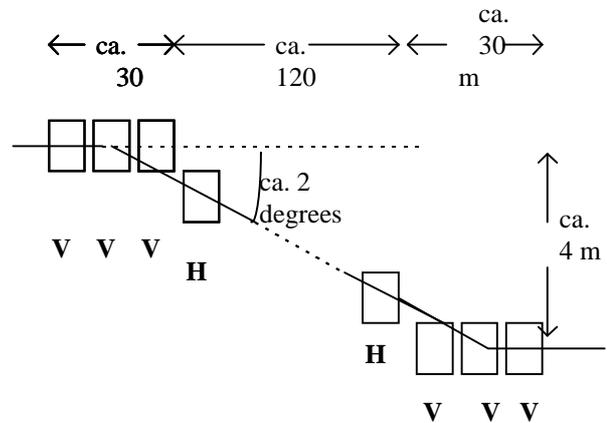

Fig. 3 Possible spin rotator for muons in the main ring. The spins are rotated 45 degrees from the vertical towards the momentum axis by the first 3 ca. 10 T, 10 m long vertically deflecting magnets. The spin is afterwards rotated by 180 degrees around the vertical axis by 12 normal bendings (spin rotation around the vertical direction) tilted by ca. 2 degrees followed by a 30 m long vertical bend into the opposite direction. The spin rotator requires an extra space of ca. 60 m.

Fig.4 shows a possible scenario for the arrangements of the spin rotators in the ring. The first spin rotator rotates the spin from the vertical (upward) into the longitudinal direction. After passing the interaction region and the second spin rotator, the spin aims in the vertical

(downward) direction. When the particle returns to the interaction region after one revolution its spin aims into the opposite direction compared to the first pass: the spin direction changes from revolution to revolution.

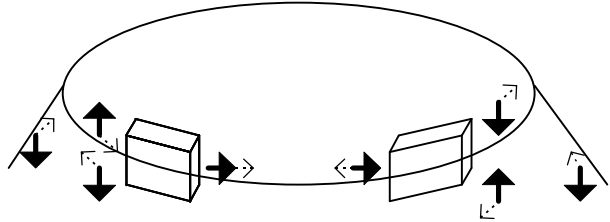

Fig. 4 shows a possible scenario for arranging the spin rotators in order to obtain varying helicity directions from interaction to interaction.

A similar statement is true for the opposite particle. As a result the helicities of the colliding bunches can be altered from revolution to revolution.

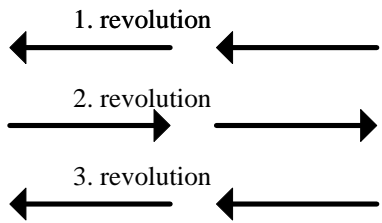

The injected beam has to be added to the beam in such a way that the spins of the muons add up. In other words the injection has to be performed every second revolution.

## 6 POLARIMETRY

The muons decay into positrons or electrons

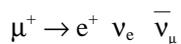

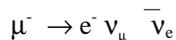

The distribution of the electrons and the positrons is commonly used as a polarimeter reaction for muons.
The lifetime for a non-relativistic muon is 2.2 μsec, γ is ca. $10^4$ and therefore the life time is 0.22 sec. One bunch contains $10^{12}$ particles. Per meter ca. $1.7 \times 10^4$ muons decay per second.
In the rest system of the $\mu^+$ the positron is mainly emitted in the forward direction as a consequence of parity violation. The decay probability is

$$W(\theta) = 1 + a_0 \cos\theta$$

where θ is the angle between the spin direction and the positron trajectory. $a_0$ depends on the energy and is 1/3 when all positron energies are taken into account. In the ring the spins aim in the vertical direction. The positrons are mainly emitted into the spin direction. In the lab frame the preferred emission angle θ becomes therefore $1/\gamma$ or $10^{-4}$ rad. This is a reasonable angle for a polarimeter.

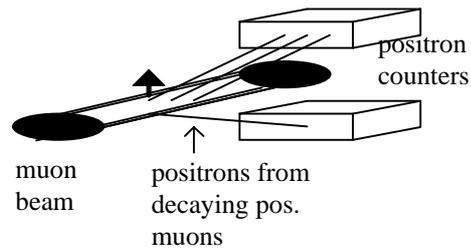

Fig. 8 A possible layout for a high energy muon polarimeter in the main ring. The asymmetry of the positrons is measured.

The asymmetry could be as high as 2 to 1 (depending on the energy selection of the particles and the acceptance of the polarimeter). It should be possible to measure polarization in a few seconds with an accuracy of 1%.